\documentclass[12pt,epsbox]{article}
\usepackage[dvips]{graphicx}
\usepackage{amsmath}

\baselineskip=\normalbaselineskip

\newcommand{\resection}[1]
 {\setcounter{equation}{0}\section{\large{#1}}}

\renewcommand{\thefootnote}{\fnsymbol{footnote}}
\newcommand{\bel}[1]{\begin{equation}\label{#1}}
\newcommand{\bal}[1]{\begin{eqnarray}\label{#1}}

\newcommand{\be}{\begin{equation}}
\newcommand{\ee}{\end{equation}}
\newcommand{\ba}{\begin{eqnarray}}
\newcommand{\ea}{\end{eqnarray}}
\newcommand{\nn}{\nonumber \\}

\newcommand{\NP}[3]{Nucl. Phys. {\bf #1} {(#2)} {#3}}
\newcommand{\PL}[3]{Phys. Lett. {\bf #1} {(#2)} {#3}}
\newcommand{\IJMP}[3]{Int. J. Mod. Phys. {\bf #1} {(#2)} {#3}}
\newcommand{\PR}[3]{Phys. Rev. {\bf #1} {(#2)} {#3}}

\newcommand{\bZ}{{\bf Z}}
\newcommand{\hg}{\widehat{g}}

\newcommand{\eq}[1]{(\ref{#1})}

\newcommand{\bS}{\mbox{\boldmath $S$}}

\renewcommand{\bZ}{\mbox{\boldmath $Z$}}

\newcommand{\hR}{\widehat{R}}

\newcommand{\tQ}{\widetilde{Q}}

\begin{document}
\setcounter{page}{0}
\begin{flushright}
\parbox{40mm}{%
YITP-01-83 \\
{\tt hep-th/0112037} \\
December 2001}

\end{flushright}

\vfill

\begin{center}
{\large{\bf Holographic Renormalization Group Structure \\ 
in Higher-Derivative Gravity}}
\end{center}

\vfill

\begin{center}
{\sc Masafumi Fukuma
\footnote{E-mail: {\tt fukuma@yukawa.kyoto-u.ac.jp}}
and
\sc So Matsuura
\footnote{E-mail: {\tt matsu@yukawa.kyoto-u.ac.jp}}
}\\[2em]

{\sl Yukawa Institute for Theoretical Physics, \\
      Kyoto University, Kyoto 606-8502, Japan } \\

\end{center}

\vfill

\begin{center}
ABSTRACT
\end{center}

\begin{quote}

\small{%
Classical higher-derivative gravity is investigated 
in the context of the holographic renormalization group (RG). 
We parametrize the Euclidean time such that  
one step of time evolution in $(d+1)$-dimensional bulk gravity 
can be directly interpreted as that of block spin transformation 
of the $d$-dimensional boundary field theory.
This parametrization simplifies 
the analysis of the holographic RG structure in gravity systems, 
and conformal fixed points are always described by AdS geometry. 
We find that higher-derivative gravity generically 
induces extra degrees of freedom 
which acquire huge mass around stable fixed points 
and thus are coupled to highly irrelevant operators at the boundary. 
In the particular case of pure $R^2$-gravity, 
we show that some region of the coefficients of curvature-squared terms 
allows us to have two fixed points (one is multicritical) 
which are connected by a kink solution. 
We further extend our analysis to Minkowski time 
to investigate a model of expanding universe 
described by the action with curvature-squared terms 
and positive cosmological constant, 
and show that, 
in any dimensionality but four, 
one can have a classical solution  
which describes time evolution from 
a de Sitter geometry to another de Sitter geometry,
along which the Hubble parameter changes drastically.
}
\end{quote}
\vfill

%
\renewcommand{\thefootnote}{\arabic{footnote}}
\setcounter{footnote}{0}
\addtocounter{page}{1}

\resection{Introduction}

The AdS/CFT correspondence states, in its simplest form, 
that $(d+1)$-dimensional (super)gravity in an AdS background 
describes a $d$-dimensional CFT at the boundary 
\cite{M}\cite{GKP}\cite{W;holography}.
(For a review, see \cite{review}.)
One of the most important aspects of this correspondence is 
that it gives us a scheme 
to investigate the renormalization group (RG) structure of 
the $d$-dimensional field theory 
\cite{SW}\cite{E.T.A}\cite{AG}\cite{HRG}%
\cite{GPPZ}\cite{GPPZ-2}\cite{PS}\cite{BK}%
\cite{ST}\cite{DFGK}.
In this scheme, the {\it holographic RG}, 
the radial coordinate of the $(d+1)$-dimensional manifold 
is identified with
the RG parameter of the corresponding boundary field theory,
and a classical trajectory of bulk fields is interpreted as 
an RG flow of the corresponding coupling constants 
in the $d$-dimensional field theory.
As an example, the Weyl anomaly of a four-dimensional field theory 
is calculated using the holographic RG scheme 
and exactly reproduces the large $N$ limit of the Weyl anomaly 
of the four dimensional ${\mathcal{N}}=4$ $SU(N)$ super 
Yang-Mills theory when supergravity 
comes from type IIB supergravity 
on AdS${}_5\times S^5$ \cite{HS;weyl}. 
For a field theory in any dimensionality, there is a systematic 
formulation of the holographic RG using the Hamilton-Jacobi 
equation of gravity systems
\cite{dVV}\cite{FMS}\cite{FS} (see also \cite{vector}\cite{FS;fermi}%
\cite{KMM}\cite{ads9}).

Classical Einstein gravity discussed above 
is actually the low energy limit of a string theory, 
and an important issue is 
whether this correspondence can be extended to the level of strings
\cite{AK}\cite{APTY}\cite{NO}\cite{BGN}\cite{BC}\cite{FMS2}. 
In \cite{FMS2}, it was discussed that the AdS/CFT correspondence 
does hold even when $\alpha'$ corrections are taken into account, 
where $\alpha'$ is the square of the string length.  
The gravity system considered in \cite{FMS2} is $R^2$-gravity 
whose Lagrangian density contains curvature squared terms 
which would appear after integrating over massive string 
excitation modes 
(such higher-derivative interactions also appear for matter fields).  
In general, a higher-derivative system%
\footnote{
See \cite{NK} which also investigates higher-derivative systems
in the context of string theory.}
with the Lagrangian 
$L(q,\dot{q},\ddot{q})$ can be treated in the Hamilton formalism 
by introducing a new independent variable $Q$ which equals $\dot{q}$
classically. 
(We call this new variable the {\it higher-derivative mode}.)
Thus the Hamiltonian for this system is a function of $(q,Q)$ and their 
conjugate momenta, $(p,P)$.
It was pointed out \cite{FMS2} 
that one can establish the AdS/CFT correspondence 
in higher-dimensional gravity 
if we take the mixed boundary conditions 
which set the Dirichlet boundary conditions for the light mode $q$ 
and the Neumann boundary conditions for the higher-derivative mode 
$Q$ ({\em i.e.}, $P=0$ at the boundary). 
As a check of this proposal, the Weyl anomaly was calculated 
for the $R^2$-gravity system 
which is AdS/CFT dual to the ${\cal N}\!=\!2$ superconformal field theory 
in four dimensions,%
\footnote%
{%
The gravity system is given by 
IIB supergravity on $AdS_5\times S^5/\bZ_2$ \cite{N=2CFT}.
The action contains an $R^2$-term, reflecting open-string excitations.
} 
and the obtained result reproduced that of \cite{NO} and \cite{BGN} 
which is  consistent with the field theoretical calculation 
\cite{Duff;Weyl}.
A brief review of classical mechanics of higher-derivative 
systems is given in Appendix A.
(For a review of higher-derivative gravity, see, e.g., \cite{HDG}.)%


The main aim of the present paper is to further clarify 
the holographic RG structure in higher-derivative gravity,
by investigating its classical solutions with the following steps. 
We first give a parametrization of the Euclidean time  
such that its evolution can be directly interpreted 
as change of the unit length of the $d$-dimensional equal time slice,  
and we call the parametrization the {\it block spin gauge}.
With the use of this gauge, 
we then investigate 
(1) a higher-derivative pure gravity system and also 
(2) a system of a scalar field 
with higher-derivative interaction in Einstein gravity. 
For both systems, some region of the coefficients of 
the higher-derivative terms allows us to have 
a stable AdS solution, 
around which the higher-derivative mode acquires huge mass 
and thus is coupled to a highly irrelevant operator at the boundary. 
In the other region of the coefficients, 
we show that any AdS solution becomes unstable 
and the higher-derivative mode in the AdS background becomes tachyonic 
with mass squared far below the unitarity bound, 
so that the holographic RG interpretation is not applicable.
We also show, in the pure gravity case, that there are two AdS
solutions in a certain region of the coefficients 
and there is also a solution which interpolates these two AdS solutions.
In the context of the holographic RG, this means 
that there are two fixed points in the phase diagram of 
the $d$-dimensional field theory, 
and that the solution which connects them corresponds to 
an RG flow from a multicritical point to another fixed point.

The organization of this paper is as follows. 
In \S 2 we introduce the block spin gauge.
In \S 3 we investigate 
a higher-derivative pure gravity system, 
and then in \S 4 we investigate a system of a scalar field 
with higher-derivative interaction in Einstein gravity. 
In \S 5, we extend our analysis to higher-derivative gravity 
with Minkowski time 
and investigate a model of expanding universe 
with positive cosmological constant.
There, we show that one can have a solution for which 
a de Sitter space-time flows to another de Sitter space-time
and the Hubble parameter changes drastically. 
\S 6 is devoted to a conclusion and a discussion about the meaning 
of the mixed boundary conditions proposed in \cite{FMS2}.

\resection{Block Spin Gauge}

In this section we introduce a gauge in which (Euclidean) time
evolution in a $(d+1)$-dimensional manifold 
is directly 
regarded as change of the unit length in the $d$-dimensional
equal time slice. 
Although this gauge restricts class of the geometry one can consider, 
it is actually enough for investigating the holographic RG structure 
in higher-derivative gravity. 

We start by recalling the ADM decomposition which parametrizes 
a $(d+1)$-dimensional metric with Euclidean signature:
\begin{align}
 ds^2 &= \hg_{\mu\nu}\,dX^{\mu}dX^{\nu} \nn
 &= N(x,\tau)^2 d\tau^2 + 
 g_{ij}(x,\tau)(dx^i+\lambda^i d\tau)(dx^j+\lambda^jd\tau),
 \label{ADM}
\end{align}
where $X^\mu=(x^i,\tau)$ with $i=1,\cdots d$, and $N$ and 
$\lambda^i$ are the lapse and the shift function, respectively.
In what follows, we exclusively consider the metric with $d$-dimensional 
Poincar\'{e} invariance by setting 
$g_{ij} = e^{-2q(\tau)}\delta_{ij}$,
$N=N(\tau)$ and $\lambda^i=0$: 
\ba
 ds^2 = N(\tau)^2 d\tau^2 + e^{-2q(\tau)} \delta_{ij} dx^idx^j.
 \label{metric}
\ea
For this metric, the unit length in the $d$-dimensional equal time slice 
at $\tau$ is given by $e^{q(\tau)}$. 

We shall consider two kinds of gauge fixing (or parametrization of time). 
One is the temporal gauge which is obtained by setting $N(\tau)=1$: 
\ba
 ds^2=d\tau^2+e^{-2q(\tau)}\delta_{ij}dx^idx^j.
 \label{temporal}
\ea
The other is a gauge fixing that can be made only when 
the condition 
\ba
 \frac{dq(\tau)}{d\tau} > 0 \qquad (-\infty < \tau < \infty) 
 \label{BScondition}
\ea
is satisfied. 
Then $q$ can be regarded as a new time coordinate, 
and we call this parametrization the {\it block spin gauge}.%
\footnote%
{
In this gauge, the unit length in the 
$d$-dimensional equal time slice at $t$ is given by $a(t)=a_0e^{t}$
with a positive constant $a_0$. 
If we consider the time evolution 
$t \to t+\delta t$, the unit length changes as $a \to e^{\delta t}a$, 
in other words, one step of time evolution directly describes 
that of block spin transformation of the $d$-dimensional field theory.
}
By writing $q(\tau)$ as $t$, 
the metric in this gauge is expressed as%
\footnote{
This form of metric sometimes appears in literature
(see, e.g., \cite{NO2}).}
\ba
 ds^2 = Q(t)^{-2}dt^2 + e^{-2t}\delta_{ij}dx^idx^j.
 \label{BSmetric}
\ea
Since two parametrizations of time (temporal and block spin) 
are related as 
\ba
 t=q(\tau)
\ea
together with the condition \eq{BScondition}, 
the coefficient $Q(t)$ is given by 
\ba
 Q(t)={dq(\tau)\over d\tau}\bigg|_{\tau=q^{-1}(t)}(>0). 
 \label{BSconditionQ}
\ea 
Note that constant $Q~(\equiv 1/l)$ gives the AdS metric of radius $l$,
\ba
 ds^2&=&d\tau^2 + e^{-2\tau/l}\,dx_i^2\quad({\rm temporal~gauge})\nn
  &=&l^2 dt^2 + e^{-2t}\,dx_i^2\quad({\rm block~spin~gauge}),
\ea
with the boundary at $\tau=-\infty$ (or $t=-\infty$).

Here we show that the condition (\ref{BScondition}) sets 
a restriction on possible geometry, 
by solving Einstein equation both in the temporal and block spin gauge.
In the temporal gauge, the Einstein-Hilbert action
\ba
 \bS_E = \int_{M_{d+1}} d^{d+1}X \sqrt{\hg}\left[2\Lambda-\hR\right], 
 \label{Einstein}
\ea
becomes 
\ba
 \bS_E = -d(d-1){\mathcal{V}}_d\int d\tau e^{-dq(\tau)}
 \left(\dot{q}(\tau)^2+\frac{1}{l^2}\right),
\ea
up to total derivative.
Here we parametrized the cosmological constant as $\Lambda = -d(d-1)/2l^2$, 
and $\mathcal{V}_d$ is the volume of the $d$-dimensional space.
The general classical solutions for this action are
\ba
 \frac{dq}{d\tau}= \frac{1}{l}\,\frac{1-Ce^{d\tau/l}}{1+Ce^{d\tau/l}}
 \qquad (C \ge 0).
\ea
This shows that geometry with nonvanishing, finite $C$ 
($C \neq 0$ or $\infty$) may not be described  
in the block spin gauge 
since $\dot{q}$ vanishes at $\tau=-\frac{l}{d}\ln{C}$, 
breaking the condition (\ref{BSmetric}). 
In fact, in the block spin gauge (\ref{BSmetric}),
the action (\ref{Einstein}) becomes
\ba
 \bS_{E} = -d(d-1){\mathcal{V}}_d\int dt e^{-dt}\left(
 \frac{1}{l^2Q}+Q\right), 
\ea
which readily gives the classical solution as 
\ba
 Q(t) = \pm \frac{1}{l}.
\ea
This actually reproduces only the AdS solution in the temporal gauge 
with $C=0$ or $\infty$. 


\resection{Higher-Derivative Pure Gravity in the Block Spin Gauge}

In this section we investigate classical $R^2$-gravity 
in the block spin gauge, 
and give a holographic RG interpretation to higher-derivative modes.
A brief review of classical mechanics of higher-derivative systems 
is given in Appendix A.


The action of pure $R^2$-gravity in a $(d+1)$-dimensional manifold 
$M_{d+1}$ with boundary $\Sigma_d$ is generally given by
\begin{align}
 \bS = &\int_{M_{d+1}}d^{d+1}X\sqrt{\hg}\left(2\Lambda-\hR
  -a\hR^2 -b\hR_{\mu\nu}^2 -c\hR_{\mu\nu\rho\sigma}^2\right) \nn
 &+\int_{\Sigma_d}d^d x \sqrt{g}\left(
  2K+x_1\,RK+x_2\,R_{ij}K^{ij}+x_3\,K^3 
  +x_4\,KK_{ij}^2+x_5\,K_{ij}^3\right),
 \label{PG:action}
\end{align}
with some given constants $a,b,c$.
Here $K_{ij}$ is the extrinsic curvature of $\Sigma_d$ given by
\ba
 K_{ij}={1\over 2N}\left(\dot{g}_{ij}
  -\nabla_i\lambda_j-\nabla_j\lambda_i\right) \qquad 
  \biggl(\cdot \equiv \frac{d}{dt}\biggr),
\ea
and $K=g^{ij}K_{ij}$. 
$\nabla_i$ and $R_{ijkl}$ are, respectively, the covariant derivative 
and the Riemann tensor defined by $g_{ij}$ in the ADM decomposition 
(\ref{ADM}). 
The first terms in the boundary terms in (\ref{PG:action}) 
is the one for Einstein gravity given in \cite{GH} and 
the remaining terms are the most general ones 
which are invariant under the $(d+1)$-dimensional diffeomorphism 
which does not change the position of the boundary.
For details, see \cite{FMS2}. 
(Another discussion of boundary terms in higher-derivative 
gravity can be found in \cite{Mye} and \cite{NO;boundary}.)

Substituting the block spin gauge metric (\ref{BSmetric}) into the
action (\ref{PG:action}), we obtain
\ba
 \bS[Q(t)] = {\mathcal{V}}_d \int_{t_0}^\infty dt\,\, L(Q,\dot{Q}),
\ea
where
\ba
 L(Q,\dot{Q}) &\!=\!&e^{-dt}\left(
  \frac{2\Lambda}{Q}-d(d-1)Q-\frac{A}{2}Q\dot{Q}^2+BQ^3\right) \nn
  &&+\left[\frac{4d}{3}\bigl(d(d+1)a+db+2c\bigr)
          +d\left(d^2x_3+dx_4+x_5\right)\right]
   \frac{d}{dt}\left(e^{-dt}Q^3\right),
  \label{PG:Lagrangian}
\ea
with
\ba
 A=2d\bigl(4da+(d+1)b+4d\bigr), \quad
 B=\frac{d(d-3)}{3}\bigl(d(d+1)a+db+2c\bigr).\label{AB}
\ea
We have set $t$ to run from $t_0$ to $\infty$. 
The Lagrangian \eq{PG:Lagrangian} gives the Euler-Lagrange equation 
for $Q$ as 
\ba
 Q\ddot{Q}+{1\over2}\dot{Q}^2-dQ\dot{Q}={1\over A}\biggl(
 {2\Lambda\over Q^2}+d(d-1)-3BQ^2\biggr).
 \label{PG:EOM} 
\ea
The classical action $S$ is obtained by substituting into $\bS$ 
the classical solution $Q(t)$ 
with the boundary condition $Q(t_0)=Q_0$ 
and the regularity of $Q(t)$ in the limit $t\to\infty$, 
and will be a function of the boundary value, 
$\bS[Q(t)] \equiv S(Q_0,t_0)$.  

In the holographic RG, this classical action would be interpreted 
as the bare action of a $d$-dimensional field theory 
with the bare coupling $Q_0$ at the UV cutoff $\Lambda=\exp (-t_0)$ 
\cite{GKP}\cite{W;holography}\cite{SW}. 
Thus, the strategy of our analysis is as follows.
We first find the solutions that converge to $Q\!=\!{\rm const.}$ 
as $t\to\infty$ in order to have a finite classical action. 
We then examine the stability of the solution 
to read off the form of general classical solutions.
Since the solution $Q\!=\!{\rm const.}$ gives AdS geometry, 
the fluctuation of $Q$ around the solution is regarded 
as describing the motion of the higher-derivative mode 
in the AdS background, 
which will lead to a holographic RG interpretation 
of the higher-derivative mode.

Following the above strategy, 
we first look for AdS solutions ({\em i.e.}, $Q(t)={\rm
const.}$).
By parametrizing the cosmological constant as
\ba
 \Lambda = -{d(d-1)\over 2l^2} + {3B\over 2l^4}, 
\ea
the equation of motion (\ref{PG:EOM}) gives two AdS solutions,
\begin{equation}
 Q^2 = 
  \begin{cases}
  \displaystyle{~~~~~~{1\over l^2}} &\displaystyle{\equiv \,{1\over l_1^2}}\,\,, 
   \vspace{2mm}\\
  \displaystyle{{d(d-1)\over 3B}-{1\over l^2}} 
   &\displaystyle{\equiv\, {1\over l_2^2}}\,\,,
   \end{cases}
 \label{AdS_sln}
\end{equation}
where the solution $Q=1/l_2$ exists only when $B>0$.%
\footnote%
{
We consider only the case $Q>0$ because of the condition 
(\ref{BScondition}).
}
They have radii $l_i\,\,(i=1,2)$, respectively, 
and we call them AdS${}^{(i)}\,\,(i=1,2)$. 
We assume that one can take the limit $a,b,c\to 0$ smoothly, 
in which the system reduces to Einstein gravity 
on AdS of radius $l$. 
We also assume that this AdS gravity comes from the low-energy limit 
of a string theory, 
so that its radius $l_1=l$ should be sufficiently larger 
than the string length.
On the other hand, the AdS${^{(2)}}$ solution, if it exists,
appears only when the higher-derivative terms are taken into account. 
As the coefficient of the higher-derivative terms are thought to 
stem from string excitations, their coefficients $a,b,c$ (and so $A,B$) 
are ${\mathcal{O}}(\alpha')$. 
Thus the radius of the AdS${^{(2)}}$ is of the order 
of the string length as can be seen from 
the solution (\ref{AdS_sln}).

Next we examine the perturbation of classical solutions 
around (\ref{AdS_sln}), writing 
\ba
 Q(t) = {1\over l_i} + X_i(t). 
\ea
The equation of motion (\ref{PG:EOM}) is then linearized as 
\ba
 \ddot{X}_i - d\dot{X}_i -l_i^2m_i^2 X_i =0,
 \label{PG:linear}
\ea
with
\ba
 m_i^2 \equiv -{2\over A}\left(2\Lambda l_i^2+ {3B\over l_i^2}\right).
 \label{mass2}
\ea
The equation (\ref{PG:linear}) is nothing but the equation of motion for
a scalar field with mass squared $m_i^2$ 
in the background of the AdS$_{d+1}$ geometry, 
$ds^2=l_i^2dt^2+e^{-2t}\sum_k dx_k^2$ (block spin gauge), 
and the general solution is given by a linear combination of
\begin{equation}
 f_i^{\pm}(t) \equiv 
  \exp\left[\left({d\over 2}\pm\sqrt{{d^2\over 4}+l_i^2m_i^2}\right)t
  \right].
 \label{PG:linear_sln} 
\end{equation}
Here $l_i^2m_i^2$ can be easily calculated from (\ref{AdS_sln}) and 
(\ref{mass2}) as
\begin{equation}
 \begin{cases}
  \displaystyle{~l_1^2m_1^2\!} 
   &\displaystyle{=\, {2\over A}\left(d(d-1)l^2-6B\right)}\,\,, \vspace{2mm}\\
  \displaystyle{~l_2^2m_2^2\!} 
  &\displaystyle{=\, -\,{6B\over A}\cdot{d(d-1)l^2-6B \over d(d-1)l^2-3B}}\,\,.
 \end{cases}
 \label{l2m2}
\end{equation}
In the following, we investigate these solutions
both for $i=1,2$, 
to understand the behavior of general classical solutions: 

\noindent
\underline{\bf perturbation around AdS${}^{(1)}$}

\noindent
From (\ref{PG:linear_sln}) and (\ref{l2m2}), the behavior of $f_1^{\pm}(t)$
depends on the signature of $A$.
For $A>0$, recalling $A$ is ${\mathcal{O}}(\alpha')$, 
$f_1^+(t)$ grows and $f_1^-(t)$ dumps very rapidly.
On the other hand, for $A<0$, the value in the square root in  
(\ref{PG:linear_sln}) becomes negative,
thus both $f_1^{\pm}(t)$ grow as $e^{dt/2}$ 
being oscillating rapidly.

\noindent
\underline{\bf perturbation around AdS${}^{(2)}$}

\noindent
We assume $B>0$ because, as mentioned before, 
AdS$^{(2)}$ exist only in that region.
For $A>0$, both of $f_2^{\pm}(t)$ grow exponentially because 
$l_2^2m_2^2<0$.
On the other hand, for $A<0$, $f_2^{+}(t)$ grows and 
$f_2^{-}(t)$ dumps exponentially.

\begin{figure}[h]
\begin{center}
\includegraphics{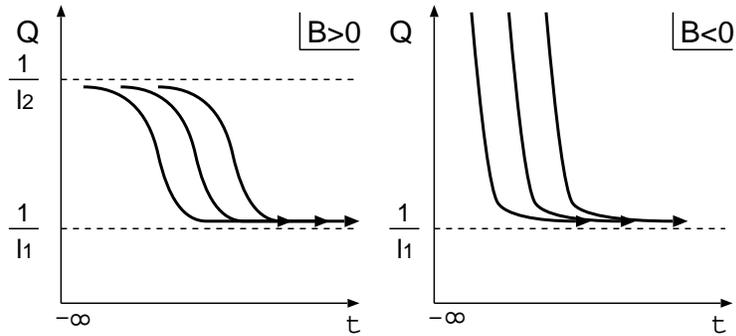}
\end{center}
\caption{\footnotesize{Classical solutions $Q(t)$ for $A>0$.}}
\label{PGflow1}
\end{figure}

\begin{figure}[h]
\begin{center}
\includegraphics{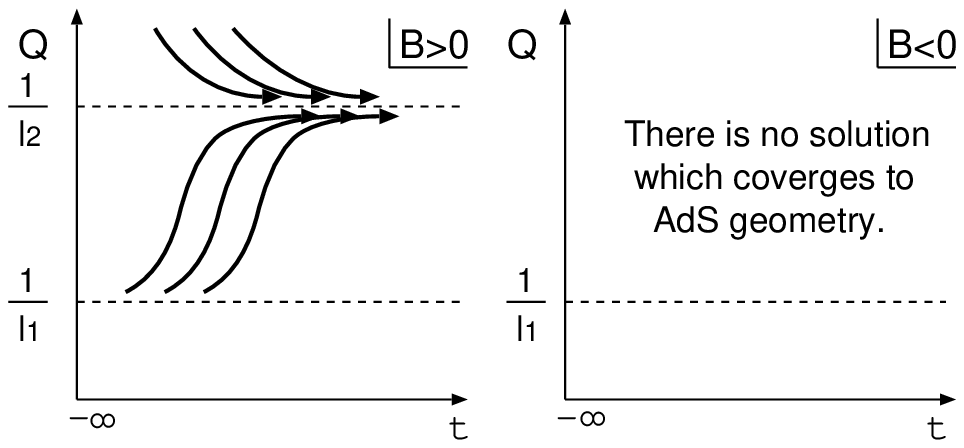}
\end{center}
\caption{\footnotesize{Classical solutions $Q(t)$ for $A<0$.}}
\label{PGflow2}
\end{figure}

\begin{figure}[h]
\begin{center}
\includegraphics[scale=.6]{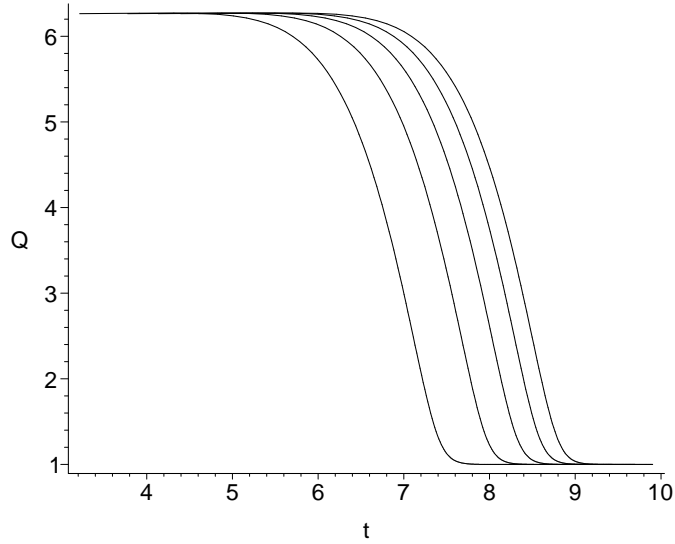}
\end{center}
\caption{\footnotesize{Result of the numerical calculation of classical solutions 
$Q(t)$ for the values
$d=4$, $A=0.1$, $B=0.1$ and $l=1$ ($1/l_1=1$ and $1/l_2=6.24$).}} 
\label{numerical}
\end{figure}

Besides, as we explained before, 
the solution which are of interest to us is such a solution 
that converges to either AdS$^{(1)}$ or AdS$^{(2)}$ as $t\to\infty$, 
satisfying the condition that $Q(t)$ be positive for all region of $t$ 
[see (\ref{BSconditionQ})]. 
After all, we can see that
the classical solutions behave as in
Fig.\,\ref{PGflow1} and Fig.\,\ref{PGflow2}.
The numerical calculation with the proper boundary condition at $t=+\infty$ 
actually exhibits these figures and shows that the branch $f_i^-(t)$ 
is selected around $Q=1/l_i$. 
The result of the numerical calculation for $A>0$ and $B>0$ is 
shown in Fig.\,{\ref{numerical}}.

Now we give a holographic RG interpretation to the above results. 
We first consider the AdS$^{(1)}$ solution. 
Eq.\ (\ref{PG:linear}) expresses the equation of motion of a scalar field 
in the AdS background of radius $l$, with mass squared given by 
\begin{align}
 m_1^2 &= -{2\over A}\left(2\Lambda l^2+ {3B\over l^2}\right) \nn
       &= {2\over A}\left(d(d-1)-{6B\over l^2}\right).
\end{align}
Thus for $A>0$, the higher-derivative mode $Q$ is interpreted as 
a very massive scalar mode, 
and thus is coupled to a highly irrelevant operator
around the fixed point 
since its scaling dimension is given by 
\cite{GKP}\cite{W;holography}
\ba
 \Delta = {d\over2}+\sqrt{{d^4\over4}+l^2m_1^2} \gg d.
\ea
This can also be understood from Fig.\,\ref{PGflow1} 
which shows a rapid convergence of the RG flow 
to the fixed point $Q(t)=1/l$.
On the other hand, for $A<0$, the mass squared of the higher-derivative
mode is far below the unitary bound $-d^2/4l^2$ 
for a scalar mode in the AdS$^{(1)}$ geometry \cite{W;holography}, 
and the scaling dimension becomes complex. 
Thus, in this case, the higher-derivative mode makes 
the AdS$^{(1)}$ geometry unstable, 
and a holographic RG interpretation cannot be given to such solution.

We then consider the AdS$^{(2)}$. 
For $A>0$ and $B>0$ in Fig.\,\ref{PGflow1}, 
one can find that classical trajectories 
begin from AdS$^{(2)}$ to AdS$^{(1)}$.
In the context of the holographic RG, 
this means that the AdS$^{(2)}$ solution $Q(t)=1/l_2$ 
corresponds to a multicritical point 
in the phase diagram of the boundary field theory.
From (\ref{AdS_sln}) and (\ref{mass2}), 
the mass squared of the mode $Q$ around the
AdS$^{(2)}$ can be calculated to be  
\ba
 m_2^2=-{2\over A}\left(d(d-1)-{6B\over l^2}\right),
\ea
and if this mass squared is above the unitarity bound,
\ba
 l_2^2m_2^2=-{6B\over A}{d(d-1)l^2-6B \over d(d-1)l^2-3B}
 > -{d^2 \over 4},
\ea
the scaling dimension of the corresponding operator is given by 
\begin{align}
 \Delta &= {d\over 2}+\sqrt{{d^2\over 4}+l_2^2m_2^2} \nn
        &\cong {d\over 2}+\sqrt{{d^2\over 4}-{6B\over A}}.
\end{align}
For example, we consider the case where $d=4$, $a=b=0$ and $c>0$.%
\footnote%
{
This includes IIB supergravity on 
AdS${}_5\times S^5/{\mathbf{Z}_2}$
which is AdS/CFT dual to ${\mathcal{N}}=2$ USp(N) SYM$_4$
\cite{N=2CFT}\cite{BGN}.
} 
In this case, $A=32c>0$ and $B=8c/3>0$, and thus the scaling
dimension of $Q$ around the AdS$^{(2)}$ is $\Delta\cong 2+\sqrt{7/2}$.
It would be interesting to investigate which conformal field theory
describes this fixed point. 

We conclude this section with a comment on the $c$-theorem. 
In the block spin gauge, the function $Q^{d-1}(t)$ can be regarded 
as the $c$-function of the $d$-dimensional field theory
\cite{HRG}. 
Fig.\ 1 shows that it increases when $A>0$, 
but this does not contradict what the $c$-theorem says 
because in this case, the kinetic term of $Q(t)$ in the bulk action 
has a negative sign [see \eq{PG:Lagrangian}].


\resection{Scalar Field with Higher-Derivative Interaction 
in Einstein Gravity}

In this section, we consider a scalar field
with higher-derivative interaction in Einstein gravity.

To simplify the discussion below, we consider the action 
\ba
 \bS=\int_{M_{d+1}}d^{d+1}X\sqrt{\hg}\left[
  V(\phi)-\hR+{1\over2}\hg^{\mu\nu}\partial_\mu\phi\partial_\nu\phi
  +{c\over2}\left(\widehat{\nabla}^2\phi\right)^2
 \right] 
  +2\int_{\Sigma_d}d^dxK,
 \label{S:action}
\ea
where $\widehat{\nabla}$ is the covariant derivative defined by 
$\hg_{\mu\nu}$, and $c$ is a given 
small constant of the order of $\alpha'$.
Substituting the block spin gauge metric (\ref{BSmetric}) into
(\ref{S:action}), $\bS$ becomes
\begin{align}
 \bS &= {\mathcal{V}}_d\int_{t_0}L(\phi,\dot{\phi},\ddot{\phi};Q) \nn
     &= {\mathcal{V}}_d\int_{t_0}e^{-dt}\left\{
 {1\over Q}V(\phi)-d(d-1)Q+{Q\over2}\dot{\phi}^2
 +{c\over2}e^{2dt}Q
 \left[\left(e^{-dt}Q\dot{\phi}\right)^{\!\displaystyle{\cdot}}\,\right]^2 
         \right\}.
 \label{S:action2}
\end{align}
As the Lagrangian contains $\ddot{\phi}$, it is convenient to treat
this system in the Hamilton formalism \cite{FMS2}. 
Following the procedure given in Appendix A, 
we introduce a Lagrange multiplier $\pi$ 
and rewrite the action in the following equivalent form:
\ba
 \bS = {\mathcal{V}}_d\int_{t_0}dt\left[
  \pi\left(\dot{\phi}-e^{dt}{\Phi\over Q}\right)
 +e^{-dt}\left({1\over Q}V(\phi)-d(d-1)Q+{Q\over2}\dot{\phi}^2\right)
 +{c\over 2}e^{dt}Q\dot{\Phi}^2
 \right].
\ea
Then, making the Legendre transformation from $\dot{\Phi}$ 
to the conjugate momentum 
\ba
 \Pi = c\,e^{dt} Q \dot{\Phi},
 \label{Phi_momentum}
\ea
we further rewrite the action into the first order form:
\ba
 \bS = {\mathcal{V}}_d\int_{t_0}dt\left[
 \pi\dot{\phi}+\Pi\dot{\Phi}-H(\phi,\Phi,\pi,\Pi;Q)
 \right],
 \label{S:firstorder}
\ea
where
\ba
 H(\phi,\Phi,\pi,\Pi;Q)=d(d-1)e^{-dt}Q+{1\over Q}\left[
 {e^{-dt}\over 2c}\Pi^2+
 e^{dt}\pi\Phi-e^{-dt}V(\phi)-{e^{dt}\over2}\Phi^2
 \right].
 \label{hamiltonian}
\ea
In (\ref{S:firstorder}), $Q$ appears without time derivative, 
thus it can be easily solved to be
\ba
 Q^2(\phi,\Phi,\pi,\Pi)={1\over d(d-1)}\left[
{1\over 2c}\Pi^2-V(\phi)
 +e^{2dt}\left(\pi\Phi-{1\over2}\Phi^2\right)
\right],
\label{S:Q}
\ea
and substituting this into the Hamiltonian (\ref{hamiltonian}), we
obtain the final form of the Hamiltonian:
\ba
 H(\phi,\Phi,\pi,\Pi)=2d(d-1)e^{-dt}Q(\phi,\Phi,\pi,\Pi).
 \label{S:hamiltonian}
\ea
The Hamilton equation is given by
\begin{equation}
 Q\dot{\phi}=e^{dt}\Phi,\quad
 Q\dot{\Phi}={e^{-dt}\over c}\Pi,\quad
 Q\dot{\pi} =e^{-dt}V'(\phi),\quad
 Q\dot{\Pi} =e^{dt}\left(\Phi-\pi\right).
 \label{S:Hamiltoneq}
\end{equation}

\begin{figure}[h]
\begin{center}
\includegraphics{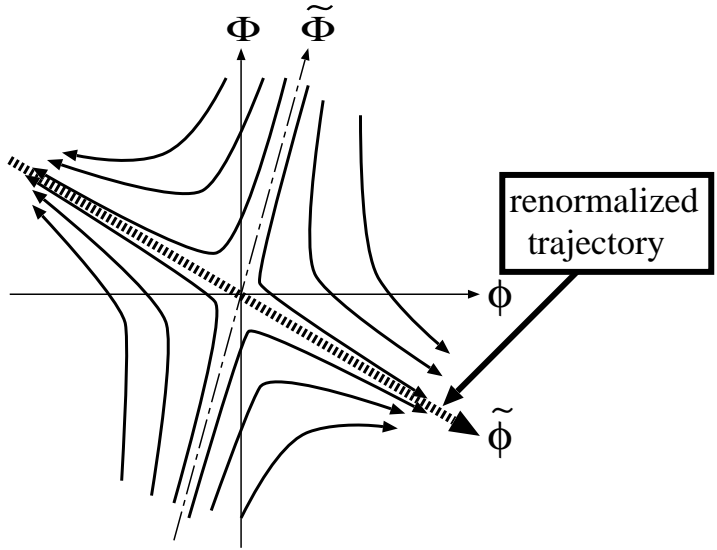}
\end{center}
\caption{\footnotesize{The RG flow of the coupling constants $(\phi,\Phi)$
near the fixed point $\phi=0$ and $\Phi=0$.}}
\label{scalar_flow}
\end{figure}

As in the pure gravity case, we first look for 
the AdS solution which is given by $Q={\rm const.}$
If we set
\ba 
 V(\phi) \equiv -{d(d-1)\over l^2}+{\mu^2\over2}\phi^2,
\ea
the AdS solution which satisfies (\ref{S:Q}) and (\ref{S:Hamiltoneq}) 
is given by
\ba
 Q=\frac{1}{l},\quad
 \phi=\Phi=\pi=\Pi=0.
 \label{S:AdS_sln}
\ea
We then expand the Hamilton equation (\ref{S:Hamiltoneq}) around 
the AdS solution (\ref{S:AdS_sln}) up to first order in variables:
\ba
 {1\over l}\dot{\phi}=e^{dt}\Phi,\quad 
 {1\over l}\dot{\Phi}={e^{-dt}\over c}\Pi,\quad
 {1\over l}\dot{\pi} =e^{-dt}\mu^2\phi,\quad 
 {1\over l}\dot{\Pi} =e^{dt}\left(\Phi-\pi\right).
 \label{Heq_lin}
\ea
This can be easily solved by performing the canonical transformation 
\cite{FMS2} 
\ba
\begin{pmatrix}
 \phi \\ \Phi \\ \pi \\ \Pi
\end{pmatrix}
= a_1
\begin{pmatrix}
 1    & 0            & 0    & e^{dt}/ M \\
 0    & e^{-dt}M     & 1    & 0               \\
 0    & e^{-dt}cm^2M & cM^2 & 0               \\
 cm^2 & 0            & 0    & e^{dt}cM        
\end{pmatrix}
\begin{pmatrix}
 \widetilde{\phi} \\ \widetilde{\Phi} \\ \widetilde{\pi} \\ \widetilde{\Pi}
\end{pmatrix},
\label{canonical}
\ea
with
\ba
 a_1^2\equiv{1\over\sqrt{1-4c\mu^2}}, \quad
 M^2\equiv{1\over 2c}\left(1+\sqrt{1-4c\mu^2}\right), \quad
  m^2\equiv{1\over 2c}\left(1-\sqrt{1-4c\mu^2}\right).
\ea
Then, the linearized Hamilton equation (\ref{Heq_lin}) is decomposed 
into two sets of independent equations,
\ba
 \begin{cases}
  \dot{\widetilde{\phi}}&=le^{dt}\widetilde{\pi}, \\
  \dot{\widetilde{\pi}}&=-lm^2e^{-dt}\widetilde{\phi},
 \end{cases}\qquad
 \begin{cases}
  \dot{\widetilde{\Phi}}&=le^{dt}\widetilde{\Pi}, \\
  \dot{\widetilde{\Pi}}&=-lM^2e^{-dt}\widetilde{\Phi},
 \end{cases}
\ea
which are equivalent to 
\begin{align}
 \ddot{\widetilde{\phi}}-d\dot{\widetilde{\phi}}-l^2m^2\widetilde{\phi}&=0,
 \\
 \ddot{\widetilde{\Phi}}-d\dot{\widetilde{\Phi}}-l^2M^2\widetilde{\Phi}&=0,
\end{align}
respectively.%
\footnote%
{
When we add the higher-derivative term $\bigl(\widehat{\nabla}^2\phi\bigr)^2$
to the action, the scalar mode is not $\phi$ but $\widetilde{\phi}$,
thus the mass of the observable field is not $\mu$ but $m$.
} 
These are nothing but the equation of motion of two scalar fields 
with mass squared $m^2$ and $M^2$, respectively, 
in the AdS background $Q=1/l$.
In particular, $\widetilde{\Phi}$ acquires large mass when $c>0$ 
since its mass squared $M^2$ becomes 
$\sim 1/c \sim 1/\alpha' \gg m^2$.
Thus the bulk scalar field $\widetilde{\Phi}$ is coupled to  
a highly irrelevant operator at the boundary.
If we assume that $\widetilde{\phi}$ is a relevant coupling, 
{\it i.e.} $-d^2/4l^2 < m^2 < 0$, 
then the RG flow near the fixed point, $\phi=\Phi=0$, will converges 
rapidly to the renormalized trajectory given by $\widetilde{\phi}=0$ 
[see Fig.\,\ref{scalar_flow}].
On the other hand, when $c<0$, the mass squared of the scalar mode 
$\widetilde{\Phi}$ is far below the unitarity bound, thus 
the AdS geometry becomes unstable.
In this case, as in the pure gravity case with $A<0,B<0$, 
the holographic
RG interpretation of the higher-derivative system is not possible.


\resection{Application to a Model of Universe 
with Positive Cosmological Constant}

In this section, we apply our analysis of higher-derivative pure gravity 
to systems of Lorentzian gravity with positive cosmological constant.
There, as classical solutions, one can have de Sitter solutions 
instead of AdS solutions.
We shall see that, in a certain region of coefficients of 
higher-derivative terms, there are two de Sitter solutions as well as  
a kink solution which interpolates these two de Sitter geometries.

We consider the following action of higher-derivative pure gravity 
in a $(d+1)$-dimensional Lorentzian Manifold:
\ba
 \bS = \int d^{d+1}X \sqrt{-\hg}\biggl(
 -2\Lambda+\hR-a\hR^2-b\hR_{\mu\nu}^2-c\hR_{\mu\nu\rho\sigma}^2
 \biggr).
 \label{dS:action}
\ea
Our discussion is completely parallel to the one given in \S 3.
We take the block spin gauge metric
\ba
 ds^2 = -{1\over Q^2}dt^2+e^{2t}\delta_{ij}dx^idx^j,
 \label{dS:BSmetric}
\ea
where we flipped the sign of the exponent to describe 
the expanding universe. 
If $Q=1/l=$const., (\ref{dS:BSmetric}) expresses de Sitter 
space-time of radius $l$.
With the metric (\ref{dS:BSmetric}), the action (\ref{dS:action}) 
becomes 
\ba
 \bS={\mathcal{V}}_d\int dt e^{dt}\left[
 -{2\Lambda\over Q}-d(d-1)Q-{A\over2}Q\dot{Q}^2+BQ^3
 \right],
\ea
where $A$ and $B$ are again given by (\ref{AB}).
This action gives the equation of motion for $Q$,
\ba
 Q\ddot{Q}+{1\over2}\dot{Q}^2+dQ\dot{Q}=
 -{1\over A}\left(
 {2\Lambda\over Q^2}-d(d-1)+3BQ^2
 \right), 
 \label{dS:EOM}
\ea
which is nothing but (\ref{PG:EOM}) if we make a change there as 
$\Lambda\to -\Lambda$ and $t\to -t$.
By parametrizing the cosmological constant as
\ba
 \Lambda={d(d-1)\over 2l^2}-{3B\over l^4},
\ea
the de Sitter solutions are obtained from (\ref{dS:EOM}) as
\ba
Q^2=
 \begin{cases}
  \displaystyle{~~~~~~~{1\over l^2}}\! 
   &\displaystyle{\equiv \,{1\over l_1^2}}\,\,, \vspace{2mm}\\
  \displaystyle{{d(d-1)\over 3B}-{1\over l^2}}\!
   &\displaystyle{\equiv\, {1\over l_2^2}} \,\,,
 \end{cases}
 \label{dSsln}
\ea
where the solution $Q=1/l_2$ exists only when $B>0$.
We call the solution $Q=1/l_i$ the dS$^{(i)}\,\,(i=1,2)$ solution,
respectively.

\begin{figure}[h]
\begin{center}
\includegraphics{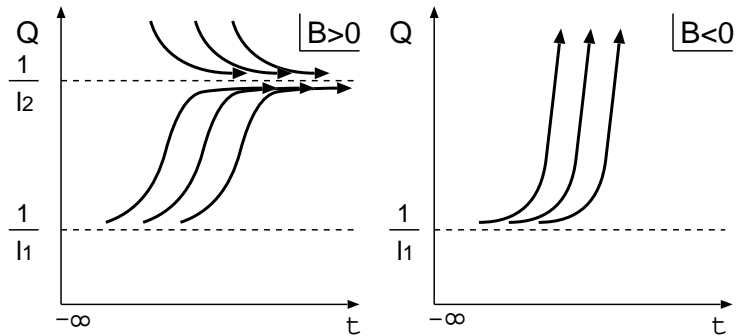}
\end{center}
\caption{\footnotesize{Classical solutions $Q(t)$ for $A>0$. The dS$^{(1)}$ solution 
is unstable, and the space-time converges to the dS$^{(2)}$ geometry 
if it exists. }}
\label{dS_to_dS1}
\end{figure}

\begin{figure}[h]
\begin{center}
\includegraphics{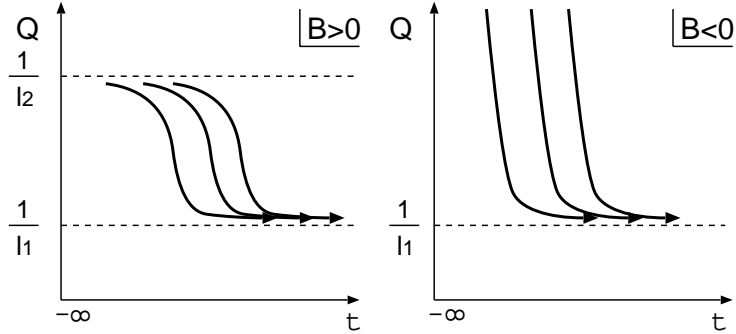}
\end{center}
\caption{\footnotesize{Classical solutions $Q(t)$ for $A<0$. The dS$^{(1)}$ solution 
is now stable, and thus the space-time converges to the dS$^{(1)}$ geometry.
If the solution dS$^{(2)}$ exists, there are solutions which describe 
time evolution from dS$^{(2)}$ to dS$^{(1)}$.}}
\label{dS_to_dS2}
\end{figure}

As we did in \S 3, we next examine the perturbation of solutions 
around these de Sitter solutions.
By writing $Q(t)$ as
\ba
 Q(t)={1\over l_i} + X_i(t),
\ea
the equation of motion (\ref{dS:EOM}) is linearized as
\ba
 \ddot{X}_i + d\dot{X}_i - \lambda_i X_i = 0,
 \label{dS:linearEOM}
\ea
where
\ba
 \begin{cases}
  \displaystyle{\,\lambda_1} &
   \displaystyle{\!\!=\, {2\over A}\left(d(d-1)l^2-6B\right)}\,\,,\vspace{2mm} \\
  \displaystyle{\,\lambda_2} &
   \displaystyle{\!\!=\,-\,{6B\over A}\cdot
   {d(d-1)l^2-6B\over d(d-1)l^2-3B}}\,\,.
  \end{cases}
 \label{lambda}
\ea
This equation is actually the time reversal of the linearized equation 
in the AdS case [see \eq{PG:linear}, \eq{l2m2}], 
and thus we readily find from Fig.\,{\ref{PGflow1}} 
and Fig.\,{\ref{PGflow2}}
that the general classical solutions behave 
as in Fig.\,{\ref{dS_to_dS1}} and Fig.\,{\ref{dS_to_dS2}}.%
\footnote{
Actually, there exist solutions which converge to the 
unstable de Sitter geometry. 
However, we ignored them
in Fig.\,{\ref{dS_to_dS1}} and Fig.\,{\ref{dS_to_dS2}} 
because such solutions form a measure-zero subspace 
in the space of classical solutions.}
Note that we now can have a meaningful solution when $A<0,\,B<0$, 
since we no longer need to restrict our consideration 
to the systems with finite classical action. 

The interesting case is when $B>0$. 
Then there is a solution which describes time evolution 
of space-time from a de Sitter geometry to another de Sitter geometry.
Since the Hubble parameter is defined by $H(\tau)=\dot{R}(\tau)/ R(\tau)$ 
for a metric 
$
 ds_{d+1}^2 = -d\tau^2 + R^2(\tau)ds_d^2 ,
$
one understands that the higher-derivative mode $Q$ is 
nothing but the Hubble parameter:
\ba
 H(\tau)=Q(t(\tau)).
\ea
Thus, the solutions for $B>0$ in Fig.\,{\ref{dS_to_dS1}} and 
Fig.\,{\ref{dS_to_dS2}} 
describe a universe in which the Hubble parameter changes rapidly 
from a constant to a constant.
Since we are assuming that the coefficients of the curvature squared terms 
are of the string scale, the difference between the two Hubble constants 
is magnificently large.
Such solutions can always exist in all dimensionality but four $(d=3)$
because $B\neq0$ when $d\neq3$.
The absence of such solutions in four-dimensional space-time 
might be remedied 
by coupling an extra matter field to gravity. 


\resection{Conclusion}

In this paper, we investigated higher-derivative gravity systems. 
We introduced the block spin gauge (\ref{BSmetric}) in which 
time evolution can be regarded directly 
as change of the unit length in the $d$-dimensional time slice.
We considered (1) higher-derivative pure gravity 
and also (2) a scalar field with higher-derivative interaction 
in Einstein gravity. 
We examined classical solutions in the block spin gauge  
and gave a holographic RG interpretation to the higher-derivative modes. 

We showed the existence of AdS solutions for both systems (1) and (2), 
and discussed their stability.
Under the request that the bulk fields be regular 
in the region far away from the boundary, 
we found that the stability of the AdS solutions depends on 
the values of the coefficients of higher-derivative terms.
In the region of stable AdS, the higher-derivative mode
can be interpreted as a very massive scalar field in the AdS
background. 
Thus, in the context of the holographic RG, 
it is coupled to a highly irrelevant operator at the boundary. 
On the other hand, in the region of unstable AdS, 
the higher-derivative mode acquires 
large negative mass squared which is far below the unitarity 
bound in AdS gravity.
In this case, it is difficult to give a holographic RG interpretation.

For higher-derivative pure gravity, in particular, 
there is a region in which one can have two AdS solutions.
In that region, one can also have a kink solution which describes 
a flow from an AdS geometry to another AdS geometry.
(This is when $B>0$ in the Fig.\,\ref{PGflow1} and Fig.\,\ref{PGflow2}.)
In particular, for $A>0$ and $B>0$, the flow starts from the AdS geometry 
of much smaller radius (of the string scale). 
This describes an RG flow from a non-trivial multicritical point 
to another fixed point, the latter of which governs the universality class 
described by pure Einstein gravity. 
The appearance of such multicritical point is characteristic 
of the holographic RG for an $R^2$-gravity system.

As an application of our analysis, we investigated $(d+1)$-dimensional 
Lorentzian higher-derivative gravity 
with positive cosmological constant. 
We found that there is a solution which describes time evolution 
from a de Sitter geometry 
to another de Sitter geometry
in a certain region of the coefficients of the curvature
squared terms.
Along the solution, the value of the Hubble parameter changes
drastically.

Finally, we comment on the meaning of the mixed boundary conditions 
which were adopted in \cite{FMS2} (see also Appendix A below).
As mentioned above, the higher-derivative mode near the stable 
AdS solution is coupled to a highly irrelevant operator at the boundary, 
so the RG flow around the corresponding fixed point 
converges rapidly to the renormalized trajectory 
on which the higher-derivative mode does not flow.
We shall see that 
one can actually pick up the renormalized trajectory
by adopting the mixed boundary conditions.

In the case of pure gravity, the fixed point
is given by the solution%
\footnote%
{
We consider only the case where the AdS$^{(1)}$ is stable.
In the presence of a scalar field which describes a relevant coupling, 
this solution corresponds to the renormalized trajectory.
}
\ba
 Q={1\over l}.
 \label{RT}
\ea
On the other hand, from the Lagrangian (\ref{PG:Lagrangian}), 
the conjugate momentum for $Q$ is calculated as
\ba
 P= -Ae^{-dt}\dot{Q}+\biggl[
    4d\bigl(d(d+1)a+db+2c\bigr)+3d\bigl(d^2x_3+dx_4+x_5\bigr)\bigg]Q^2
\ea
Thus, the fixed point \eq{RT} can be picked up 
by the equation $P=0$ 
if we set the coefficients as in \cite{FMS2}:
\ba
 d^2x_3+dx_4+x_5 = -{4\over3}\bigl(d(d+1)a+db+2c\bigr).
\ea
In other words, with the use of the freedom to add total derivative terms 
to the action, 
the coefficients can be chosen such that the equation $P=0$ directly 
gives the fixed point.
Note that the total derivative terms can be interpreted 
as the generating function of a canonical transformation 
which shifts the value of the conjugate momentum. 

The situation does not change for a scalar field coupled to 
Einstein gravity with higher-derivative interaction.
When $\widetilde{\phi}$ is a relevant coupling, 
the renormalized 
trajectory is given by $\widetilde{\Phi} = 0$, which is equivalent 
to $\widetilde{\Pi}=0$.
On the other hand, from the canonical transformation (\ref{canonical}),
$\widetilde{\Pi}$ is expressed as
\ba
 \widetilde{\Pi} = \sqrt{1-4c\mu^2}\biggl(
 \Pi-cm^2\phi\biggr).
\ea
Thus, if we add the term
\ba
 {d\over dt}F(\phi,\Phi) \equiv {d\over dt}\biggl(cm^2\phi\Phi\biggr)
\ea
to the Lagrangian (\ref{S:action2}) 
(or equivalently $c m^2 \widehat{\nabla}^\mu(\phi\,\partial_\mu\phi)$ 
to the Lagrangian density), 
we can shift the conjugate momenta as
\ba
 \pi \to \pi +cm^2\Phi,\quad \Pi \to \Pi + cm^2\phi,
\ea
so that we have $\widetilde{\pi}\propto\pi$ and 
$\widetilde{\Pi}\propto\Pi$. 
This enables us to pick up the renormalized trajectory 
with the mixed boundary conditions ($\Pi=0$).

\section*{\large{Acknowledgments}}

The authors would like to thank T.\ Sakai for discussions 
and collaboration at the early stage of this work. 
They also thank T.\ Kubota, H.\ Kudoh, M.\ Ninomiya and 
S.\ Nojiri for helpful 
discussions.

\appendix

\resection{General Theory of Higher-Derivative Systems}

In this appendix, we give a brief review on classical mechanics of 
higher-derivative systems with the action
\ba
 \bS{}[q] = \int_{t_0}^{t_1} dt\,L(q,\dot{q},\ddot{q}).
 \label{HD:action}
\ea
The variational principle gives the Euler-Lagrange equation:
\ba
 0={d^2\over dt^2}\left({\partial L \over \partial \ddot{q}}\right)
   -{d\over dt}\left({\partial L \over \partial \dot{q}}\right)
   +{\partial L \over \partial q}.
 \label{HD:ELeq}
\ea

This system can also be investigated in the Hamilton formalism: 
We first introduce a Lagrange multiplier $p$ to treat $\dot{q}$ 
as a new canonical variable $Q$:
\ba
 L\left(q,\dot{q},\ddot{q}\right) \to 
 p\,(\dot{q}-Q)+L\left(q,Q,\dot{Q}\right).
\ea
We call $Q$ the {\it higher-derivative mode}.
Then, by making the Legendre transformation from $\dot{Q}$ 
to the conjugate momentum $P\equiv \partial L/\partial \dot{Q}$, 
this action can be rewritten into the first order form:
\ba
 \bS{}[q,Q,p,P]=\int_{t_0}^{t_1} dt\,
 \left[p\,\dot{q}+P\dot{Q}-H(q,Q;p,P)\right],
 \label{HD:firstorder}
\ea
with the Hamiltonian 
\ba
 H(q,Q;q,P) \equiv p\,Q+P \cdot f(q,Q,P)-L(q,Q,f(q,Q,P)).
 \label{ham} 
\ea
Here $f(q,Q,P)$ in (\ref{ham}) is obtained by solving 
$P=\partial L(q,Q,f)/\partial f=P(q,Q,f)$ in $f$.
Again by the variational principle, we obtain the Hamilton equation
\ba
 \dot{q}={\partial H\over\partial p},\quad
 \dot{Q}={\partial H\over\partial P},\quad
 \dot{p}=-{\partial H\over\partial q},\quad
 \dot{P}=-{\partial H\over\partial Q},
 \label{HD:hamiltoneq}
\ea
together with the boundary conditions
\ba
 p\,\delta q + P\delta Q =0 \quad (t=t_0,t_1).
 \label{BC} 
\ea
One can easily check that the Hamilton equation (\ref{HD:hamiltoneq})
is equivalent to the Euler-Lagrange equation (\ref{HD:ELeq}).

The boundary condition (\ref{BC}) is satisfied by the Dirichlet 
boundary conditions
\ba
 \delta q=0\,,\quad \delta Q=0 \quad (t=t_0,t_1)\,,
\ea
or the Neumann boundary conditions 
\ba
 p=0\,,\quad P=0\quad (t=t_0,t_1)\,,
\ea
for each variable $q$ and $Q$.
A choice of interest for us is to take the mixed boundary conditions,  
that is, we set the Dirichlet conditions for $q$ 
($q(t_0)=q_0$ and $q(t_1)=q_1$) 
and the Neumann conditions for $Q$ ($P(t_0)=P(t_1)=0$). 
Then, if we substitute such classical solution into the bulk action $\bS$, 
the resulting classical action is a function only of the 
boundary values of the light mode $q$; 
$\bS[q(t),Q(t),p(t),P(t)]\equiv S(q_0,t_0;q_1,t_1)$. 

In \cite{FMS2}, the mixed boundary conditions were adopted
to establish the holographic principle in higher-derivative gravity systems.
In fact, if we set the mixed boundary conditions for a bulk field $\phi(x,t)$  
as $\phi(x,t\!=\!t_a)=\phi_a(x)$ and $\Pi(x,t\!=\!t_a)=0$ 
$(a=0,1)$,%
\footnote%
{
$\Pi$ is the conjugate momentum of the higher-derivative mode 
$\Phi$ ($\sim \dot{\phi}$).
} 
and carefully choose $\phi_1(x)$ 
such that the classical action is finite in the limit $t_1\to +\infty$, 
then the classical action becomes a functional only of 
$\phi_0(x)$ and $t_0$, $S[\phi_0(x),t_0]$. 
This may be interpreted as the fixed point action 
with the bare coupling $\phi_0$ 
at the UV cutoff $\Lambda=\exp(-t_0)$, 
in the presence of an irrelevant operator 
corresponding to the higher-derivative mode of $\phi$. 
In other words, the classical solution under the mixed boundary conditions 
may describe an RG flow 
of the coupling constant along the renormalized trajectory. 
The main text of the present paper actually supports this idea.


\resection{Higher-Derivative Pure Gravity without Gauge Fixing}

In this appendix, we verify that $Q(t)$ in the block spin gauge metric 
is actually the higher-derivative mode in the sense given in Appendix A. 
We give a discussion by explicitly solving the equation of motion 
of (\ref{PG:action}) 
without assuming any particular form for the variables 
appearing in the metric (\ref{metric}).  

Substituting (\ref{metric}) into (\ref{PG:action}), we obtain 
the Lagrangian of this system:%
\footnote{
Here we ignore the boundary terms because they don't affect 
the equation of motion.} 
\begin{align}
 L\biggl(q,&{\dot{q}\over N},
  \left({\dot{q}\over N}\right)^{\!\!\displaystyle{\cdot}};N\biggr) \nn
    &=Ne^{-dq(\tau)}\left\{2\Lambda 
      -d(d-1)\left(\dot{q}(\tau)\over N\right)^2
      -{A\over2N^2}\left[\left(\dot{q}(\tau)\over
      N\right)^{\!\!\displaystyle{\cdot}}\,\right]^2 
      +B\left(\dot{q}(\tau)\over N\right)^4
 \right\},
\end{align}
where $\cdot\equiv d/d\tau$.
Following the discussion of Appendix A, we introduce a Lagrange
multiplier $p$ to set
\ba
 \tQ(\tau) = {\dot{q}(\tau)\over N}.
\ea
The Lagrangian then becomes
\ba
 L=p\left(\dot{q}-N\tQ\right) -{A\over 2N}e^{-dq}\dot{\tQ}^2
   +Ne^{-dq}\left(2\Lambda-d(d-1)\tQ^2+B\tQ^4\right).
\ea
Since $N$ is not dynamical, its classical value can be easily found to be
\ba
 N=\sqrt{A\dot{\tQ}^2\over
 2p\tQ e^{dq}-2\left(2\Lambda-d(d-1)\tQ^2+B\tQ^4\right)}.
\ea
Substituting this into the Lagrangian, we obtain the action for this system:
\ba
 \bS=\int_{\tau_0}d\tau \left\{p\dot{q}+2\sqrt{
 {A\over 2}e^{-dq}\dot{\tQ}^2
 \left[p\tQ-e^{-dq}\left(2\Lambda-d(d-1)\tQ^2+B\tQ^4\right)\right]}
 \right\}.
\ea
Now we impose the condition (\ref{BScondition}) to $q(\tau)$,
which allows us to change the integration variable from $\tau$ to $q$:
\ba
 \bS=\int_{q_0}dq \left\{p+2\sqrt{
 {A\over 2}e^{-dq}\dot{Q}^2
 \biggl[pQ-e^{-dq}\left(2\Lambda-d(d-1)Q^2+BQ^4\right)\biggr]}
 \right\},
 \label{BS:action2}
\ea
where 
\ba
 Q(q) \equiv \tQ(\tau(q)),
\ea
and $\cdot$ is now understood to represent $d/dq$.  
The action (\ref{BS:action2}) can be further simplified 
by substituting the classical value of $p$, 
and we finally obtain the action 
\ba
 \bS = \int_{q_0} dq \,\,
 e^{-dq}\left(-\frac{A}{2}\,Q\,\dot{Q}^2
  +\frac{2\Lambda}{Q}-d(d-1)Q+BQ^3\right).
\ea
This is nothing but the action \eq{PG:Lagrangian} in the block spin gauge 
if we rewrite $q$ as $t$.
Thus we can conclude that $Q(t)$ in the block spin gauge metric 
(\ref{BSmetric}) corresponds to the higher-derivative mode 
introduced in Appendix A, 
and is related to the variable $q$ in the temporal gauge ($N=1)$ as 
\ba
 Q(t) = {dq(\tau) \over d\tau}\bigg|_{\tau=q^{-1}(t)}.
\ea
Using the same procedure, we can also derive (\ref{S:hamiltonian}) 
from the temporal gauge metric (\ref{temporal}) under the condition
(\ref{BScondition}).



\begin{thebibliography}{99}
\bibitem{M}
J.\ Maldacena,
``{\it The large $N$ limit of superconformal field theories and 
supergravity},''
Adv.\ Theor.\ Math.\ Phys.\ {\bf 2} (1998) 231,
hep-th/9711200.

\bibitem{GKP}
S.\ S.\ Gubser, I.\ R.\ Klebanov and A.\ M.\ Polyakov,
``{\it Gauge Theory Correlators from Non-Critical String Theory},''
Phys.\ Lett.\ {\bf B428} (1998) 105, 
hep-th/9802109. 

\bibitem{W;holography}
E.\ Witten,
``{\it Anti De Sitter Space And Holography},''
Adv.\ Theor.\ Math.\ Phys.\ {\bf 2} (1998) 253,
hep-th/9802150.

\bibitem{review}
O.\ Aharony, S.\ S.\ Gubser, J.\ Maldacena, H.\ Ooguri and Y.\ Oz,
``{\it Large N Field Theories, String Theory and Gravity},''
hep-th/9905111, and references therein.

\bibitem{SW}
L.\ Susskind and E.\ Witten, 
``{\it The holographic bound in anti-de Sitter space},'' 
hep-th/9805114.

\bibitem{E.T.A}
E.\ T.\ Akhmedov,
``{\it A remark on the AdS/CFT correspondence and the renormalization
 group flow},''
Phys.Lett. {\bf B442} (1998) 152,
hep-th/9806217.

\bibitem{AG}
E.\ Alvarez and C.\ Gomez,
``{\it Geometric Holography, the Renormalization Group 
and the c-Theorem},''
Nucl.Phys. {\bf B541} (1999) 441,
hep-th/9807226.

\bibitem{HRG}
D.Z.\ Freedman, S.S.\ Gubser, K. Pilch and N.P.\ Warner, 
``{\it Renormalization Group Flows from Holography--Supersymmetry 
and a c-Theorem},''
hep-th/9904017.

\bibitem{GPPZ}
L.\ Girardello, M.\ Petrini, M.\ Porrati and A.\ Zaffaroni, 
``{\it Novel Local CFT and Exact Results on Perturbations of N=4 Super
	Yang Mills from AdS Dynamics},''
J.High Energy Phys. {\bf 12} (1998) 022,
hep-th/9810126.

\bibitem{GPPZ-2}
L.\ Girardello, M.\ Petrini, M.\ Porrati and A.\ Zaffaroni
``{\it The Supergravity Dual of N=1 Super Yang-Mills Theory},''
Nucl.Phys. {\bf B569} (2000) 451,
hep-th/9909047.

\bibitem{PS}
M.\ Porrati and A.\ Starinets,
{\it  ``RG Fixed Points in Supergravity Duals of 4-d Field Theory and
	Asymptotically AdS Spaces},''
Phys.Lett. {\bf B454} (1999) 77,
hep-th/9903085.


\bibitem{BK}
V.\ Balasubramanian and P.\ Kraus,
``{\it Spacetime and the Holographic Renormalization Group},''
Phys.\ Rev.\ Lett.\ {\bf 83} (1999) 3605,
hep-th/9903190.

\bibitem{ST}
K.\ Skenderis and P.\ K.\ Townsend,
{\it  ``Gravitational Stability and Renormalization-Group Flow},''
Phys.Lett. {\bf B468} (1999) 46,
hep-th/9909070.

\bibitem{DFGK}
O.\ DeWolfe, D.\ Z.\ Freedman, S.\ S.\ Gubser and A.\ Karch,
``{\it Modeling the fifth dimension with scalars and gravity},'' 
Phys.\ Rev.\ {\bf D62} (2000) 046008.
hep-th/9909134.

\bibitem{HS;weyl}
M.\ Henningson and K.\ Skenderis,
``{\it The Holographic Weyl anomaly},''
J.High Energy Phys. {\bf 07} (1998) 023,
hep-th/9806087.

\bibitem{dVV}
J.\ de Boer, E.\ Verlinde and H.\ Verlinde,
``{\it On the Holographic Renormalization Group},''
hep-th/9912012.

\bibitem{FMS}
M.\ Fukuma, S.\ Matsuura and T.\ Sakai, 
``{\it A Note on the Weyl Anomaly in the Holographic Renormalization 
Group},''
Prog.\ Theor.\ Phys.\ {\bf 104} (2000) 1089,
hep-th/0007062.

\bibitem{FS}
M.\ Fukuma and T.\ Sakai, 
``{\it Comment on Ambiguities in the Holographic Weyl Anomaly},''
Mod.\ Phys.\ Lett.\ {\bf A15} (2000) 1703,
hep-th/0007200.

\bibitem{vector}
S.\ Corley,
``{\it A Note on Holographic Ward Identities},''
Phys.\ Lett.\ {\bf B484} (2000) 141,
hep-th/0004030.

\bibitem{FS;fermi}
J.\ Kalkkinen and D.\ Martelli, 
``{\it Holographic Renormalization Group with Fermions and Form Fields},''
hep-th/0007234.

\bibitem{KMM}
J.\ Kalkkinen, D.\ Martelli and W.\ Mueck, 
``{\it Holographic Renormalisation and Anomalies},''
hep-th/0103111.

\bibitem{ads9}
S.\ Nojiri, S.\ D.\ Odintsov and S.\ Ogushi, 
``{\it Holographic renormalization group and conformal anomaly for 
AdS$_9$/CFT$_8$ correspondence },''
\PL{500}{2001}{199},
hep-th/0011182. 

\bibitem{AK}
D.\ Anselmi and A.\ Kehagias, 
``{\it Subleading Corrections and Central Charges in the AdS/CFT 
Correspondence},''
\PL{B455}{1999}{155} ,
hep-th/9812092.

\bibitem{APTY}
O.\ Aharony, J.\ Pawelczyk, S.\ Theisen and S.\ Yankielowicz,
``{\it A Note on Anomalies in the AdS/CFT correspondence},''
Phys.\ Rev.\ {\bf D60} (1999) 066001,
hep-th/9901134.

\bibitem{NO}
S.\ Nojiri and S.\ D.\ Odintsov, 
``{\it On the conformal anomaly from higher derivative gravity 
in AdS/CFT correspondence},''
\IJMP{A15}{2000}{413}
hep-th/9903033; \\
S.\ Nojiri and S.\ D.\ Odintsov, 
``{\it Finite gravitational action for higher derivative and stringy 
gravity},''
\PR{D62}{2000}{064018}
hep-th/9911152. 

\bibitem{BGN}
M.\ Blau, K.\ S.\ Narain and E.\ Gava
``{\it On Subleading Contributions  to the AdS/CFT Trace Anomaly},''
J.High Energy Phys. {\bf 9909} (1999) 018,
hep-th/9904179. 

\bibitem{BC}
A.\ Bilal and C.-S.\ Chu, 
``{\it A Note on the Chiral Anomaly in the AdS/CFT Correspondence and 
$1/N^2$ Correction},''
\NP{B562}{1999}{181}, 
hep-th/9907106. 

\bibitem{FMS2}
M.\ Fukuma, S.\ Matsuura and T.\ Sakai, 
``{\it Higher-Derivative Gravity and the AdS/CFT Correspondence},''
Prog.\ Theor.\ Phys.\ {\bf 105} (2001) 1017,
hep-th/0103187.

\bibitem{NK}
S.\ Naka, 
``{\it Difference Equation and Dual Resonance Model},''
Prog.\ Theor.\ Phys.\ {\bf 48} (1972) 1024;\\
M.\ Kato, 
``{\it Particle theories with minimum observable length and 
open string theory},''
Phys.\ Lett.\ {\bf B245} (1990) 43. 

\bibitem{N=2CFT}
A.\ Fayyazuddin and M.\ Spalinski 
``{\it Large $N$ Superconformal Gauge Theories and Supergravity
	Orientifolds},''
Nucl.Phys. {\bf B535} (1998) 219,
hep-th/9805096; \\
O.\ Aharony, A.\ Fayyazuddin and J.\ Maldacena,
``{\it The Large $N$ Limit of ${\cal N}=1,2$ Field Theories from Three
	Branes in F-theory},''
J.High Energy Phys. {\bf 9807} (1998) 013,
hep-th/9806159.


\bibitem{Duff;Weyl}
M.\ J.\ Duff,
``{\it Twenty Years of the Weyl Anomaly},''
Class.\ Quant.\ Grav.\ {\bf 11} (1994) 1387,
hep-th/9308075.


\bibitem{HDG}
D.\ G.\ Boulware, 
``{\it Quantization of Higher Derivative Theoreis of Gravity},''
in Quantum theory of gravity : essays in honor of the 60th birthday 
of Bryce S. DeWitt, \ 
ed. S.\ M.\ Christensen (Adam Hilger Ltd, Bristol, 1984), P.267. \\
S.\ W.\ Hawking and J.\ C.\ Luttrell, 
``{\it Higher Derivative in Quantum Cosmology (I). The isotropic case},''
\NP{B247}{1984}{250}.

\bibitem{NO2}
S.\ Nojiri and S.\ D.\ Odintsov,
``{\it Brane World Inflation Induced by Quantum Effects},''
Phys. Lett. {\bf B484} (2000) 119,
hep-th/0004097.

\bibitem{GH}
G.\ W.\ Gibbons and S.\ W.\ Hawking,
``{\it Action Integrals and Partition Functions in Quantum Gravity},''
Phys. Rev. {\bf D15} (1977) 2752.

\bibitem{Mye}
R.\ C.\ Myers, 
``{\it Higher-derivative gravity, surface terms, and string theory},''
Phys. Rev. {\bf D36} (1987) 392.

\bibitem{NO;boundary}
S.\ Nojiri and S.\ D.\ Odintsov, 
``{\it Brane-World Cosmology in Higher Derivative Gravity or Warped 
Compactification in the Next-to-leading Order of AdS/CFT Correspondence},''
J.High Energy Phys. {\bf 0007} (2000) 049,
hep-th/0006232, \\
S.\ Nojiri, S.\ D.\ Odintsov and S.\ Ogushi, 
``{\it Dynamical Branes from Gravitational Dual of ${\cal N}\!=\!2$ 
$Sp(N)$ Superconformal Field Theory},''
hep-th/0010004, \\ 
S.\ Nojiri, S.\ D.\ Odintsov, S.\ Ogushi,
``{\it Holographic entropy and brane FRW-dynamics from AdS black hole 
in d5 higher derivative gravity},''
hep-th/0105117.


\end{thebibliography}
\end{document}